\documentclass[pra,twocolumn,titlepage,nofootinbib,amsmath,showpacs]{revtex4}%
\usepackage{amsmath}
\usepackage{amsfonts}
\usepackage{amssymb}
\usepackage{graphicx}%
\begin{document}
\title{Non-relativistic contributions in order $\alpha^5m_\mu c^2$ to the Lamb shift in muonic hydrogen, deuterium
and helium ion}
\author{S.~G.~Karshenboim}
\email{savely.karshenboim@mpq.mpg.de} \affiliation{D.~I. Mendeleev
Institute for Metrology, St.Petersburg, 190005, Russia}
\affiliation{Max-Planck-Institut f\"ur Quantenoptik, Garching,
85748, Germany}
\author{V.~G.~Ivanov}
\affiliation{Pulkovo Observatory, St.Petersburg, 196140, Russia}
\affiliation{D.~I. Mendeleev Institute for Metrology, St.Petersburg,
190005, Russia}
\author{E.~Yu.~Korzinin}
 \affiliation{D.~I. Mendeleev
Institute for Metrology, St.Petersburg, 190005, Russia}
\author{V. A. Shelyuto}
 \affiliation{D.~I. Mendeleev
Institute for Metrology, St.Petersburg, 190005, Russia}

\newcommand{\eq}[1]{(\ref{#1})}
\begin{abstract}
Contributions to the energy levels in light muonic atoms and, in
particular, to the Lamb shift fall into a few well-distinguished
classes. The related diagrams are calculated using different
approaches. In particular, there is a specific kind of
non-relativistic contributions. Here we consider such corrections to
the Lamb shift in order $\alpha^5m_\mu$. These contributions are due
to free vacuum polarization loops as well as to various effects of
light-by-light scattering. The closed loop in the related diagrams
is an electronic one, which allows a non-relativistic consideration
of the muon. Both kinds of contributions have been known for a
while, however, the results obtained up to date are only partial ones.
We complete a calculation of the $\alpha^5m_\mu$ contributions for
muonic hydrogen. The results are also adjusted for muonic deuterium
and muonic helium ion.
\end{abstract}
\maketitle



Recent progress of the PSI experiment on the Lamb shift in muonic
hydrogen \cite{mu_lamb} has attracted interest to theory of the
Lamb shift in light muonic atoms. Their study  can provide us with
information on certain nuclear structure effects with accuracy that
is not available in any other experiment.

To obtain such data, one has to be able to separate quantum
electrodynamics (QED) effects from the nuclear structure effects,
and for this purpose an adequate QED theory providing high accuracy
is required. Contributions to the energy levels in light muonic
atoms and, in particular, to the Lamb shift fall into a few
different well-distinguished classes. A specific theory stands behind
each of them. There are corrections, the evaluation of which is
identical for hydrogen and muonic hydrogen, and corrections that are
specific for muonic atoms. The latter involve a certain part of QED,
recoil effects and effects of the finite nuclear size. An
important class of such specific contributions, which, in fact, also
include the dominant term for the Lamb shift, is due to
non-relativistic physics.

We remind that  atomic momenta in light muonic atoms $\sim Z\alpha
m_\mu\simeq 1.5 Z m_e$ are compatible with the electron mass, while
the atomic energy $\sim (Z\alpha)^2 m_\mu\simeq 0.01 Z^2 m_e$ is
much smaller than the electron mass. (The relativistic units in
which $\hbar=c=1$ are applied throughout the paper.) Such an
environment produces an important sector of corrections, which deal
with a non-relativistic bound muon, while the QED effects are
present only through the closed electron loops. Meanwhile the
Compton wave length of electron $\lambdabar_e=1/m_e$ determines the
radius of the effective interaction induced by this kind of
diagrams. The loops may be for either free-loop vacuum-polarization
(VP) effects, related to the Uehling, K\"allen-Sabry potential and
higher-order diagrams, or the to light-by-light (LbL) scattering
contributions.

\begin{figure}[thbp]
\vspace{-4pt}
\begin{center}
\resizebox{0.95\columnwidth}{!}{\includegraphics{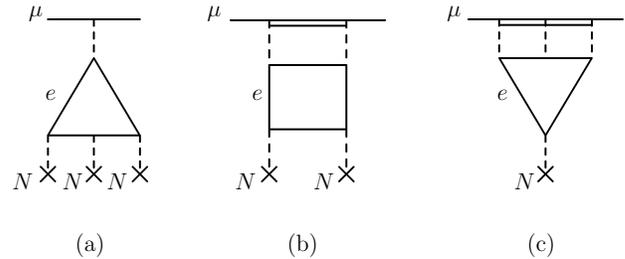}}
\end{center}
\vspace{-7pt} \caption{Characteristic diagrams for three basic
contributions of light-by-light scattering effects to the Lamb shift
in muonic hydrogen in order $\alpha^5m_\mu$. Here, $N$ stands for a
nucleus, which may be a proton, a deuteron etc. The horizontal
double line is for the muon propagator in the Coulomb field.}
\label{fig:lbl}       
\end{figure}

The VP leading term is of order $\alpha(Z\alpha)^2m_\mu$ and it has
been known for a while, while the second-order VP contribution (of
order $\alpha^2(Z\alpha)^2m_\mu$) was calculated with appropriate
accuracy for muonic hydrogen in \cite{pach} only relatively
recently.

\begin{figure}[thbp]
\vspace{-4pt}
\begin{center}
\resizebox{0.95\columnwidth}{!}{\includegraphics{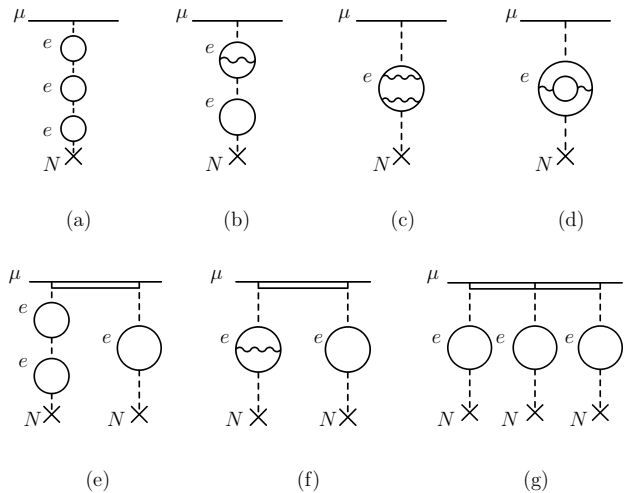}}
\end{center}
\vspace{-7pt} \caption{Characteristic diagrams for free electron-VP
contributions to the Lamb shift in muonic hydrogen in order
$\alpha^5m_\mu$.}
\label{fig:vp}       
\end{figure}

The accuracy of current and planned experiments requires a complete
theory of non-relativistic contributions to the Lamb shift in order
$\alpha^5m_\mu$. The LbL contributions are depicted in
Fig.~\ref{fig:lbl}, while the vacuum polarization diagrams are
present in Fig.~\ref{fig:vp}. Both kinds of contributions have been
known for a while, however, the results obtained up to now were only
partial ones. In particular, in the case of muonic hydrogen the
contribution in Fig.~\ref{fig:lbl}$c$ has not yet been calculated,
while there are also some questions \cite{EGSb} about applicability
of the so-called scattering approximation applied in \cite{bor_h} to
evaluate the contribution in Fig.~\ref{fig:lbl}$b$.


Certain LbL contributions have specific names. The first one in
Fig.~\ref{fig:lbl} is a so-called Wichmann-Kroll contribution and it
was calculated for muonic hydrogen with sufficient accuracy in
\cite{EGS,EGSb}. It was also reproduced in \cite{bor_h}; we also
confirm this contribution. For muonic deuterium and muonic helium-4
ion the results have been obtained in \cite{bor_d,bor_he} and we
confirm the deuterium result \cite{bor_d} and obtain a result for a
muonic helium ion
\begin{equation}\label{wk:he}
\Delta E_{\ref{fig:lbl}a}^{\rm He}=-0.0198(4)\;{\rm meV}\;,
\end{equation}
which is consistent with $-0.02\;$meV of \cite{bor_he}, but more
accurate, and strongly disagrees with $+0.135\;$meV of \cite{mart}.

In our calculation we used approximations for the Wichmann-Kroll
potential in the form
\begin{eqnarray}
V_{\rm
 WK}(r)&=&\frac{\alpha(Z\alpha)^2}{\pi}\frac{Z\alpha}{r}\times 0.361662331\nonumber\\
 &\times&\exp{\Bigl[\,0.3728079 x}\nonumber\\
 &-&\sqrt{4.416798 x^2+11.39911 x+2.906096}\,\Bigr]
 \nonumber
\end{eqnarray}
as discussed in \cite{EGSb} (see also \cite{approx}) and
 \begin{equation}
 V_{\rm
 WK}(r)=\frac{(Z\alpha)^310^{-4}}{r}\left\{
 \begin{array}{lr}
 \frac{1.528-0.489 x}{1.374 x^3+1.41 x^2+2.672 x+1}\;, & x\leq 1\\
 \\
 \frac{0.207 x^2+0.367 x-0.413}{x^6}\;,                & x>1
 \end{array}\right.\nonumber
 \end{equation}
as considered in \cite{rmp} and \cite{pach}. Here $x=m_er$. The
results are consistent. Indeed, if higher accuracy is required, one
can apply an exact expression \cite{blomquist} for $ V_{\rm WK}(r)$
as a two-dimensional integral.

The second term (Fig.~\ref{fig:lbl}$b$) is called
`virtual-Delbr\"uck-scattering contribution'. It has been calculated
for muonic hydrogen in \cite{bor_h}. The calculation was based on
\cite{scattering,rmp}, where at first a scattering approximation was
applied and subsequently a number of further approximations was
made. We remind that the scattering approximation suggests that the
external muon legs in the diagram in Fig.~\ref{fig:lbl}$b$ are
on-shell (i.e. $p^2=m_\mu^2$) and the muon propagator there is
substituted for a free one, i.e., the kinematics is exactly the same
if one calculates a related Born scattering amplitude. Since atomic
momenta $Z\alpha m_\mu$ in light muonic atoms are compatible with
the electron mass $m_e$, the validity of such an approximation is
questionable (see, e.g., the discussion in \cite{EGSb}).

We, however, have proved that the scattering approximation is
applicable within the uncertainty of order $(Z\alpha)^2m_\mu/m_e$
(in fractional units), which is at the level of about 1\% in muonic
hydrogen and deuterium and of about 4\% in muonic helium. That is
also correct for other simplifying approximations, which were made
in the calculations for this contribution in light muonic atoms
\cite{scattering,rmp,bor_h,bor_d,bor_he}. A general idea of our
evaluation is presented in Appendix~\ref{s:a}, while the details of
our evaluation are to be published elsewhere \cite{elsewhere}.

Eventually, we conclude that the uncertainty of the method applied
in \cite{bor_h,bor_d,bor_he} is substantially smaller than the
uncertainty of the related numerical evaluations for the Lamb shift
correction in muonic hydrogen \cite{bor_h}, muonic deuterium
\cite{bor_d} and muonic helium-4 ion \cite{bor_he}.

The contribution in Fig.~\ref{fig:lbl}$c$ has no specific name.
Since any other LbL contribution has one (`Wichmann-Kroll term' and
`virtual-Delbr\"uck-scattering contribution'), sometimes it is referred
to as a  `light-by-light contribution', which is somewhat
confusing.

This contribution has remained uncalculated for a while. Studying
applicability of the scattering approximation for the diagram in
Fig.~\ref{fig:lbl}$b$, we have also managed to prove
\cite{elsewhere} that this remaining contribution can be expressed
in terms of the well-known Wichmann-Kroll term
\begin{equation}\label{3:1}
\Delta E_{\ref{fig:lbl}c}=\frac{1}{Z^2}\Delta E_{\ref{fig:lbl}a}\;.
\end{equation}
The uncertainty of this identity is of order
$(Z\alpha)^2m_\mu/m_e$ (in fractional units), which is at the level
of about 1\% in muonic hydrogen and deuterium and of about 4\% in
muonic helium.

By combining our results on the uncertainty of various approximations
with the numerical results of other authors, we obtain the complete
result for all LbL contributions of Fig.~\ref{fig:lbl}. The result
is listed in the summary table (Table~\ref{t:sum}).

With identity (\ref{3:1}) proved and a possibility to obtain a
result for the Wichmann-Kroll ($\Delta E_{\ref{fig:lbl}a}$) with
high accuracy for any light muonic atom, the uncertainty in the
calculation of the complete LbL contribution now comes from the
virtual-Delbr\"uck-scattering term, which should determine the
eventual uncertainty of the non-relativistic $\alpha^5m_\mu$ term
for muonic hydrogen, deuterium and helium-ion.


Another major non-relativistic contribution in order $\alpha^5m_\mu$
is due to vacuum polarization contributions. The VP terms of this
order were studied for muonic hydrogen in \cite{vp31}. The diagrams
are depicted in Fig.~\ref{fig:vp}, which includes contributions of
the first (Fig~\ref{fig:vp}$a$--$d$), second
(Fig~\ref{fig:vp}$e,\;d$) and the third (Fig~\ref{fig:vp}$f$) order
of non-relativistic perturbation theory (NRPT).

The most complicated terms are indeed related to the first line of
Fig.~\ref{fig:vp}, however, these diagrams were cross-checked due to
their contributions to the anomalous magnetic moment of muon
\cite{baikov,vp31g} and we can rely on them.


The contributions of the second line of Fig.~\ref{fig:vp} are
specific for muonic atoms and do not correlate directly with any
calculation for the muon $g\!-\!2$. Those have been recalculated
completely independently of \cite{vp31}, as well as part of
diagrams in the first line.

We confirm the second-order terms of NRPT, while our result
\cite{vp32} for the third-order term (the last diagram in
Fig.~\ref{fig:vp}) disagrees with the one originally published in
\cite{vp31}. After a correction \cite{vp31e} their result agrees with
ours. In our calculations we used techniques developed while
investigating second-order vacuum-polarization contributions to the
hyperfine structure of muonic hydrogen \cite{muhfs}.

 \begin{table}[htbp]
 \begin{center}
 \begin{tabular}{|c|c|c|c|}
 \hline
Term& $\Delta E^{\rm H}(2p\!-\!2s)$& $\Delta E^{\rm
D}(2p\!-\!2s)$ & $\Delta E^{\rm He}(2p\!-\!2s)$\\
 & [meV]& [meV] & [meV]\\
 \hline
1st order VP & 205.007\,36       & 227.634\,67       & 1665.7729  \\
2nd order VP &  1.658\,85         & 1.838\,04         & 13.2769 \\
3rd order VP&  0.007\,52         & 0.008\,42(7)$^\star$      & 0.074(3)$^\star$ \\
LbL (Fig.~\ref{fig:lbl}) & $-0.000\,71(15)$$^\star$ & $-0.000\,73(16)$$^\star$ &  $-0.005(10)$$^\star$\\
\hline
NR total &  206.673\,02(15) & 229.480\,40(17) & 1679.119(10) \\
 \hline
 \end{tabular}
\caption{The non-relativistic QED contributions to the Lamb shift
$\Delta E(2s\!-\!2p)$ in light muonic atoms: hydrogen (H), deuterium
(D), helium-4 ion (He), which includes VP contributions of the first
($\alpha(Z\alpha)^2m_\mu$), the second ($\alpha^2(Z\alpha)^2m_\mu$)
and the third $\alpha^3(Z\alpha)^2m_\mu$ (Fig.~\ref{fig:vp}) order
as well as a complete LbL contribution. The latter is a sum of
contributions of order $\alpha(Z\alpha)^4m_\mu$
(Fig.~\ref{fig:lbl}$a$), $\alpha^2(Z\alpha)^3m_\mu$
(Fig.~\ref{fig:lbl}$b$) and $\alpha^3(Z\alpha)^2m_\mu$
(Fig.~\ref{fig:lbl}$c$). Results marked with $\star$ are obtained in
this work.\label{t:sum}}
 \end{center}
 \end{table}

The diagrams in Fig.~\ref{fig:vp} were also discussed in various
papers in the context of the Lamb shift in muonic deuterium \cite{bor_d}
and muonic helium-4 ion \cite{mart}. For this purpose a part of the
contributions was recalculated there.

Here, we reevaluated all the vacuum-polarization contributions and
the results are listed in Table~\ref{t:sum}. Comparing with the paper
mentioned we have to acknowledge that our results are not in
complete agreement with theirs.

In \cite{bor_d} only contributions of Fig.~\ref{fig:vp}$a$ and $b$
were directly calculated for muonic deuterium and the results for
them agree with ours. However, that result were not applied there but
instead the muonic-hydrogen result was `re-scaled'.
That was achieved by assuming that the result for muonic hydrogen,
presented in \cite{vp31} in the form of
\begin{equation}
\Delta
E(2p-2s)=C_3\,\left(\frac{\alpha}{\pi}\right)^3(Z\alpha)^2m_{\rm
r}\;,
\end{equation}
where $m_{\rm r}$ is the muon reduced mass, can be directly applied
for the muonic deuterium. That has not been claimed in \cite{vp31}
and is indeed incorrect
and the value of the coefficient \cite{vp32,vp31e}
\begin{eqnarray}\label{c3hyd}
C_3=0.118\,680(12)
\end{eqnarray}
is valid only for muonic hydrogen (cf. with $C_3=0.120\,045(12)$
from \cite{vp31}, which needs a correction \cite{vp32} as explained
above). The related values for other light muonic atoms, obtained
here,
\begin{equation}\label{obs1}
 C_3=\left\{
 \begin{array}{ll}
0.1262(11)\;, ~~~& {\rm for}~\mu {\rm D}\;,\\
0.270(17) \;, ~~~&  {\rm for}~\mu{\rm He}^+\;.
 \end{array}\right.
 \end{equation}
obviously differ from (\ref{c3hyd}).

The muonic-helium-4 paper \cite{mart} lacks a complete result and
only a part of diagrams of Fig.~\ref{fig:vp} were recalculated. Our
results are not in  fair agreement, and in particular we strongly
disagree in the contribution of Fig.~\ref{fig:vp}$e$ for muonic
helium. In the Wichmann-Kroll contribution (Fig.~\ref{fig:lbl}$a$)
we also strongly disagree with the result \cite{mart}, while we
agree with the result \cite{bor_he}, for which we obtain higher
accuracy (see (\ref{wk:he})).

Finally, we summarize in Table~\ref{t:sum} a complete theory of
non-relativistic QED contributions to the Lamb shift in muonic hydrogen,
deuterium and helium ion up to the order $\alpha^5m_\mu$ (see
\cite{EGSb} for references to the calculation of the low-order
corrections).\\

A part of this work was done during visits of  VGI, EYK, \& VAS to
Garching and they are grateful to MPQ for the hospitality. This work
was supported in part by RFBR (grants \#\# 08-02-91969  \&
08-02-13516) and DFG (grant GZ 436 RUS 113/769/0-3). The work of EYK
was also supported by the Dynasty foundation. The authors are
grateful to Randolf Pohl, Aldo Antognini, and Tobias Nebel for
stimulating discussions.\\


{\bf Note Added:\/} 
After submission of this work, we have
calculated the LbL contribution within a static-muon approximation
(see Appendix A). The preliminary results for $\Delta
E_{1b}$, which are 0.001\,15(1) meV for $\mu$H,
0.001\,24(1) meV for $\mu$D, and 0.0114(4) meV for $\mu$He$^+$, are
somewhat below the former results \cite{bor_h,bor_d,bor_he}, but
still are in fair agreement with them. These more accurate results
will be reported in detail in a future publication \cite{eprint2010}.

\appendix

\section{On approximations for the LbL contribution\label{s:a}}

We have proven a kind of theorem \cite{elsewhere} that the diagrams
in Fig.~\ref{fig:lbl} can be calculated  in light muonic atoms (for
simplicity we consider further muonic hydrogen) within the
static-muon approximation, in which the complete muon-line factor
shrinks to
\[
{\cal F}({\bf q})=\int {\frac{d^3\!p}{(2\pi)^3}}\;\Psi^*({\bf
p})\;\Psi({\bf p}+{\bf q})\;,
\]
where ${\bf q}$ is the total momentum transfer to the muon line,
$\Psi$ is the wave function and the error is of the order of
$(Z\alpha)^2m_\mu/m_e$.


The scattering approximation \cite{bor_h,bor_d} agrees with the
static-muon approximation within the same uncertainty.

\begin{figure}[thbp]
\begin{center}
\resizebox{1.0\columnwidth}{!}{\includegraphics{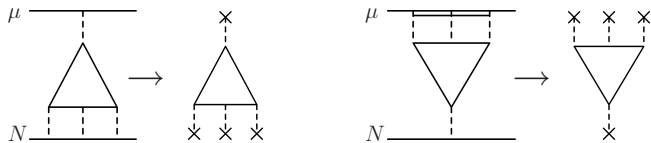}}
\end{center}
\caption{Reduction of diagrams with free/bound fermion propagators
to static-fermion diagrams (i.e. to the static-proton approximation
($=$ the external-field approximation) (left) and to the static-muon
approximation (right).}
\label{fig:eq}       
\end{figure}

The proof of the `theorem' is to be presented elsewhere
\cite{elsewhere} and here we explain its main idea. The proton and
muon are both non-relativistic (NR) particles. Their descriptions are
very similar.

Let us first compare the related diagrams with free NR fermion
propagators (cf. Fig.~\ref{fig:eq}). The expressions for the muon
and proton lines are identical. Apparently, one can approximate the
proton line within the external field approach or, which is the
same, by a static proton. To transform the complete NR expressions
to the static-proton case, we have to neglect the proton kinetic
energy in the proton propagators. Once that is done, after a chain
of identities we should arrive at the external-field approximation.

A reason to neglect the energy is the fact that a characteristic
energy, related to a particle of the mass $M$, is $E_M\sim
\gamma^2/M$, where $\gamma=Z\alpha m_\mu$ is a characteristic atomic
momentum. One can prove that we can expand using small parameters
$E_M/m_e$ and $E_M/\gamma$. However, in muonic hydrogen they are of
the same order since $\gamma\sim m_e$. The parameter $E_M/m_e$ (and
the related error) differs indeed for a muon and proton, but it is
small for both.

Thus as long as the muon propagator is a free one there is no
difference in proving that we can apply the static-muon
approximation and the static-proton approximation (see
Fig.~\ref{fig:eq}).

Meantime, in reality the situation for a muon and a proton is
somewhat different. The muon characteristic momentum is of the same
order as the electron mass $\gamma\sim m_e$, and we should treat it
as a bound one.

The NR Coulomb Green function of a muon includes
\[
G_C(E,{\bf p},{\bf p}^\prime)=i\sum_\lambda \frac{\vert\lambda({\bf
p})\rangle   \langle\lambda({\bf p}^\prime)\vert}{E-E_\lambda+i0}\;,
\]
a summation over all intermediate states $\lambda$ of continuous and
discrete spectrum, involving energy of the intermediates. The
characteristic energy of an intermediate state is indeed of order of
magnitude of the atomic bound energy $E_\lambda\sim\gamma^2/m_\mu$
and we can neglect it, as we already did in the case of the free
propagators. After that, the sum over intermediates shrinks to the
unity operator and the Coulomb propagator becomes equal to a free
one with the kinetic-energy term ${\bf p}^2/2m_\mu$ neglected
\[
i\sum_\lambda \frac{\vert\lambda\rangle
\langle\lambda\vert}{E-E_\lambda+i0}\to i\sum_\lambda
\frac{\vert\lambda\rangle
\langle\lambda\vert}{E+i0}=\frac{i}{E+i0}\;.
\]

\end{document}